\documentclass[3p,times,twocolumn]{elsarticle}

\usepackage{ecrc}

\volume{00}

\firstpage{1}

\journalname{Nuclear and Particle Physics Proceedings}

\runauth{}

\jid{nppp}

\jnltitlelogo{Nuclear and Particle Physics Proceedings}

\usepackage{amssymb}

\usepackage[figuresright]{rotating}
\usepackage{amsmath}
\usepackage{amssymb}

\usepackage{graphicx}
\usepackage{dcolumn}
\usepackage{bm}
\usepackage{subfig}
\usepackage{placeins}
\usepackage{breqn}

\usepackage{titlesec}
\titlespacing*{\section}{0pt}{0.1\baselineskip}{0.1\baselineskip}
\titlespacing*{\subsection}{0pt}{0.1\baselineskip}{0.1\baselineskip}
\titlespacing*{\abstract}{0pt}{0.1\baselineskip}{0.1\baselineskip}

\newcommand{\fg}[1]{Fig.\hspace{0.7mm}\ref{#1}}
\newcommand{\eq}[1]{Eqn.\hspace{0.7mm}\ref{#1}}

\makeatletter
\g@addto@macro\normalsize{%
  \setlength\abovedisplayskip{2pt}
  \setlength\belowdisplayskip{2pt}
  \setlength\abovedisplayshortskip{2pt}
  \setlength\belowdisplayshortskip{2pt}
}
\makeatother

\begin{document}

\begin{frontmatter}



\dochead{}

\title{AdS/CFT predictions for azimuthal and momentum correlations of $b\bar{b}$ pairs in heavy ion collisions}


\author[uct]{R. Hambrock}
\ead{roberthambrock@gmail.com}
\author[uct]{W. A. Horowitz}
\ead{wa.horowitz@uct.ac.za}

\address[uct]{Department of Physics, University of Cape Town, Private Bag X3, Rondebosch 7701, South Africa}

\begin{abstract}
  We use an energy loss model sensitive to thermal fluctuations \cite{arXiv:1501.04693} to compute the azimuthal and momentum correlations of $b\bar{b}$ pairs traversing a strongly coupled plasma from Pb+Pb collisions at LHC ($\sqrt{s}=2.76\text{TeV}$). The azimuthal correlations are compared with those from perturbative QCD based simulations \cite{arXiv:1305.3823}. When restricted to leading order production processes, we find that the strongly coupled correlations of high transverse momentum pairs ($>4\text{GeV}$) are broadened less efficiently than the corresponding weak coupling based correlations, while low transverse momentum pairs (1-4GeV) are broadened with similar efficiency, but with an order of magnitude more particles ending up in this momentum class.

  The strong coupling momentum correlations we compute account for initial correlations and reveal that the particle pairs suppressed from initially high momenta to the low momentum domain do not suffice to explain the stark difference to the weak coupling results in momentum correlations for 1-4GeV.
  From this, we conclude that $b\bar{b}$ pairs are more likely to stay correlated in momentum when propagating through a strongly coupled plasma than a weakly coupled one.
\end{abstract}


\end{frontmatter}


\section{\label{sec:Introduction}Introduction}
The quark gluon plasma is of great interest since it represents our first case study of the emergent physics of the non-abelian gauge theory QCD. A key step in understanding this state of matter is identifying its relevant coupling strength. The perturbative techniques of QCD are only adequate in a weakly coupled plasma, with calculations for strongly coupled plasmas constrained to methods like AdS/CFT-based approaches or Resonance Scattering.
Both weak and strong coupling based approaches have had their respective successes in the past. For instance, experimental $R^\pi_{AA}$ measurements show surprisingly consistent agreement with predicions from pQCD based models \cite{arXiv:1210.8330}, while AdS/CFT based calculations have fared strongly by predicting a global lower bound on the shear viscosity-to-entropy ratio of QGP-like systems of $\frac{\eta}{s}\sim0.1$ \cite{hep-th/9602135}, when taking natural units, which is in line with hydrodynamic inferences from collider data at LHC and at RHIC \cite{arXiv:1105.3226}.
Both frameworks show qualitative agreement with $R_{AA}^D$ \cite{ arXiv:1409.7545, arXiv:1410.0692}, suggesting they are attaining sufficient maturity to investigate more differential observables.
We will argue that the momentum correlations of heavy quarks constitute a promising candidate as a differentiator between weakly and strongly coupled plasmas.

In \cite{arXiv:1305.3823}, the azimuthal correlations of heavy $q\bar{q}$ pairs in a weakly coupled plasma in Pb+Pb collisions ($\sqrt{s}=2.76\text{TeV}$) were studied, both for a model involving purely collisional energy loss and one additionally incorporating radiative corrections.
These weak coupling based azimuthal correlations provide a secondary indicator for the momentum correlations and we will compare them with computations from an AdS/CFT correspondence exploiting energy loss model sensitive to thermal fluctuations, the latter already having been introduced in \cite{arXiv:1501.04693}.
As in \cite{arXiv:1501.04693}, we will probe the spectrum of reasonable AdS/CFT based energy loss models with two plausible 't Hooft coupling constants ($\lambda_1 = 5.5$ \& $\lambda_2=12\pi\alpha_s\approx11.3$ with $\alpha_s=0.3$) with the additional requirements that, for the former, temperature and the Yang-Mills coupling are equated, while in the latter case, energy density and the coupling are equated.

The calculations will be performed for the same transverse momentum classes as in \cite{arXiv:1305.3823} and also both with leading order and next-to-leading order production processes used for the initialisation.
Additionally, we will consider momentum correlations that take initial momentum correlations into account. These will provide evidence that heavy quarks traversing a strongly coupled plasma are more likely to stay correlated in momentum than they would if inside a weakly coupled plasma.

\section{\label{sec:EnergyLossModel}Energy Loss Model}

\subsection{\label{subsec:Overview}Overview}
The following will outline our computational procedure and its background.
Subsequent to initializing the momenta of bottom quark pairs either to leading order with FONLL \cite{arXiv:1205.6344} or to next-to-leading order with aMC@NLO \cite{arXiv:1405.0301} using Herwig++ \cite{arXiv:0803.0883} for the showering, the production points of the bottom quarks are weigthed by the Glauber binary distribution \cite{arXiv:1501.04693}.
The particles are propagated through the plasma via the energy loss mechanism described in \ref{subsec:LangevinEnergyLoss} until the temperature in their local fluid cell drops below the $T_c$ threshold and hadronization was presumed to occur or $8.6\text{fm}$ had passed, being the maximum time of the VISHNU background \cite{arXiv:1105.3226}.
\subsection{\label{subsec:LangevinEnergyLoss}Langevin Energy Loss}
As shown in \cite{hep-ph/0412346}, the stochastic equation of motion for a heavy quark in the fluid's rest frame is given by
\begin{equation}
  \frac{dp_i}{dt}=-\mu p_i + F_i^L + F_i^T \label{eqn:stochastic_eom}
\end{equation}
where $F_i^L$ and $F_i^T$ are longitudinal and transverse momentum kicks with respect to the quark's direction of propagation and with $\mu$, the drag loss coefficient, being given by $\mu=\pi \sqrt{\lambda} T^2/(2M_Q)$ \cite{hep-th/0605158},
where $M_Q$ is the mass of a heavy quark in a plasma of temperature $T$ with 't Hooft coupling constant $\lambda$.
The correlations of momentum kicks are given by
\begin{align}
  \langle F_i^T(t_1)F_j^T(t_2) \rangle &= \kappa_T(\delta{ij}-\frac{\vec{p}_i\vec{p}_j}{|p|^2})g(t_2-t_1) \label{eqn:F_T}\\
  \langle F_i^L(t_1)F_j^L(t_2) \rangle &= \kappa_L\frac{p_ip_j}{|p|^2}g(t_2-t_1) \label{eqn:F_L}
\end{align}
where g is only known numerically \cite{arXiv:1501.04693} and with
\begin{align}
  \kappa_T&=\pi\sqrt{\lambda}T^3\gamma \label{eqn:kappa_T}\\
  \kappa_L&=\gamma^2\kappa_T=\pi\sqrt{\lambda}T^{5/2} \label{eqn:kappa_L}
\end{align}
As via \eq{eqn:kappa_L}, the coupling of the longitudinal fluctuations to velocity grows as $\gamma^{5/2}$, thus growing significant extremely quickly \cite{arXiv:1501.04693}. As reasoned in \cite{arXiv:1501.04693}, the fluctuations are thus important to include for finite $\lambda\sim\mathcal{O}(10)$, as is the case for our probing models, where  $\gamma_{crit}^{fluc}=\frac{M_Q^2}{4T^2}$
is lower than the speed limit on a quark, $\gamma_{crit}^{sl}=(1+\frac{2M_Q}{\sqrt{\lambda}T})^2$
where $g(0)=1$, since any kick will be fully correlated to itself
If the time scale of momentum kick correlations is small compared the time scale determined by the drag coefficient, we can model the colouring as white noise, hence treat $g$ as a Dirac delta \cite{arXiv:1501.04693}.
By virtue of our requirement $\gamma<\gamma_{crit}^{sl}$, it follows that
  $t_{corr}\mu\sim\frac{1}{2}\sqrt{\lambda}\sqrt{\gamma}\frac{T}{M_Q}<\frac{1}{2}\sqrt{\lambda}\sqrt{4M_Q^2/\lambda T^2}\frac{T}{M_Q}=1$,
and we may thus safely approximate the colouring as white noise.

The computations based on this model will be labeled "Gubser".

\subsection{\label{subsec:LangevinEnergyLossMoerman}Development on \ref{subsec:LangevinEnergyLoss}}

The problem with the energy loss mechanism described in \ref{subsec:LangevinEnergyLoss} is that since the longitudinal momentum fluctuations grow as $\gamma^{\frac{5}{2}}$, our setup breaks down for high momenta.

In \cite{arXiv:1605.09285}, the average squared distance travelled by offshell quarks in a strongly coupled is computed via the $AdS/CFT$ correspondence: as in \cite{hep-th/0612143}, we consider $s^2(t, a, d)$ of the free endpoint of a string initially at rest, where the other endpoint is fixed to a black hole horizon in $AdS_d$.
For asymptotically late times, we find
\begin{equation}
  s^2(t \gg \beta, a, d) = s_{\mbox{small}}^2(t \gg \beta, a, d) =
  \frac{(d-1)^2}{8\pi\sqrt{\lambda}}\beta(1-\frac{a}{2})
\end{equation}
where $\beta=T^{-1}$ and $a$ parametrizes between a heavy quark for $a=0$ and a light quark for $a=1$.
At late times, the motion is diffusive, thus we can extract the diffusion coefficient
\begin{equation}
  D(a, d = 5) \sim s^2(t \gg \beta, a, d = 5)/2 = 2\beta/(\pi\sqrt{\lambda})
\end{equation}
Thus, the transverse momentum fluctuations are
\begin{equation}
  \kappa_T=2T^2/D=\pi\sqrt{\lambda}T^2/\beta=\pi\sqrt{\lambda}T^3
\end{equation}
Then we attain $\hat{q}$ via
\begin{equation}
  \hat{q}=\frac{\langle p_\perp(t)^2 \rangle}{\lambda}
  \approx \frac{2\kappa_Tt}{\lambda} = \frac{\pi T^3 t}{\sqrt{\lambda}}
\end{equation}
The computations based on this model will be labeled "MH".



  %

\section{Leading Order Correlations}
\label{sec:Leading Order Correlations}

\begin{figure}
  \subfloat{
  \includegraphics[width=.4\textwidth]{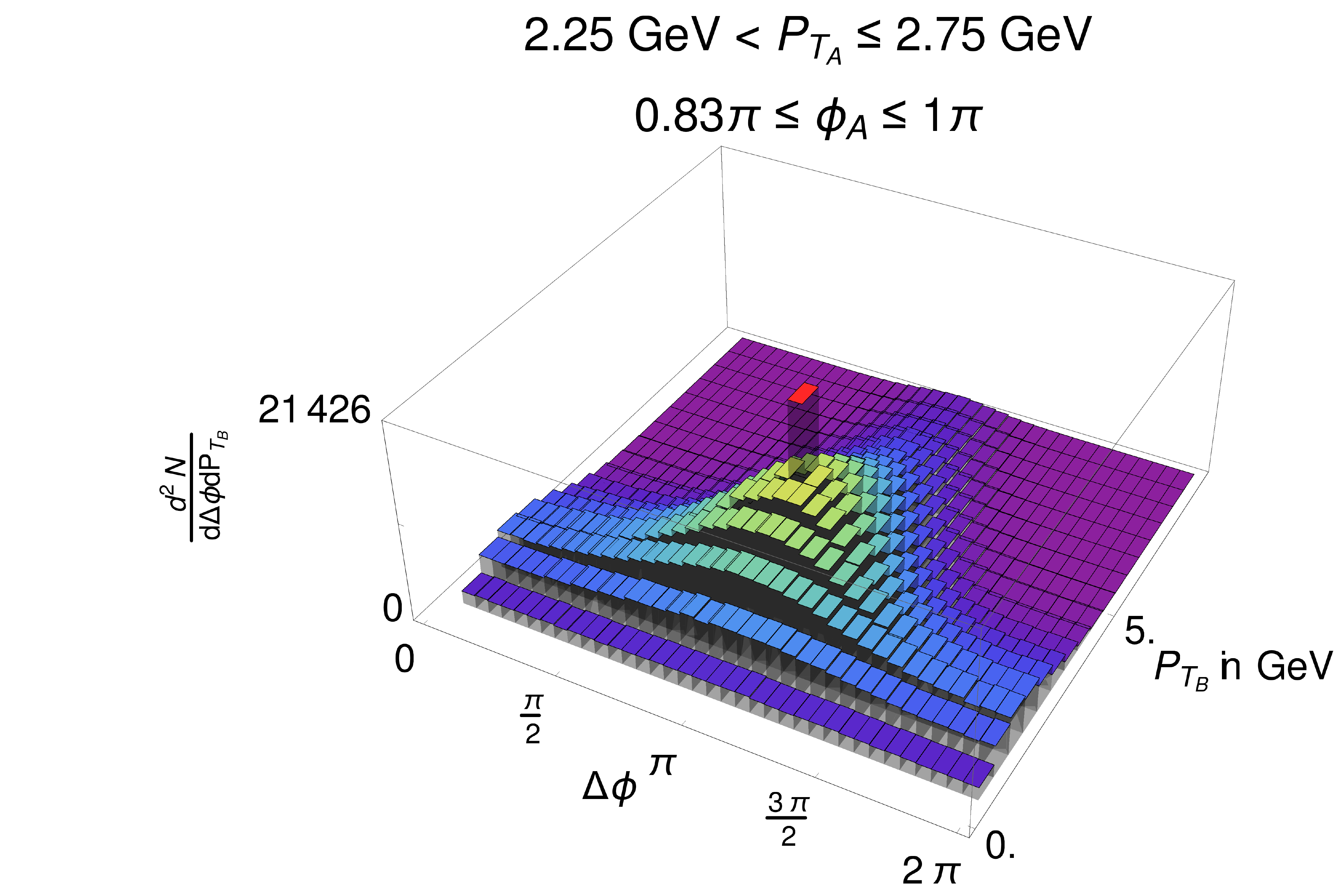}
  }
  \hfill
  \subfloat{
  \includegraphics[width=.4\textwidth]{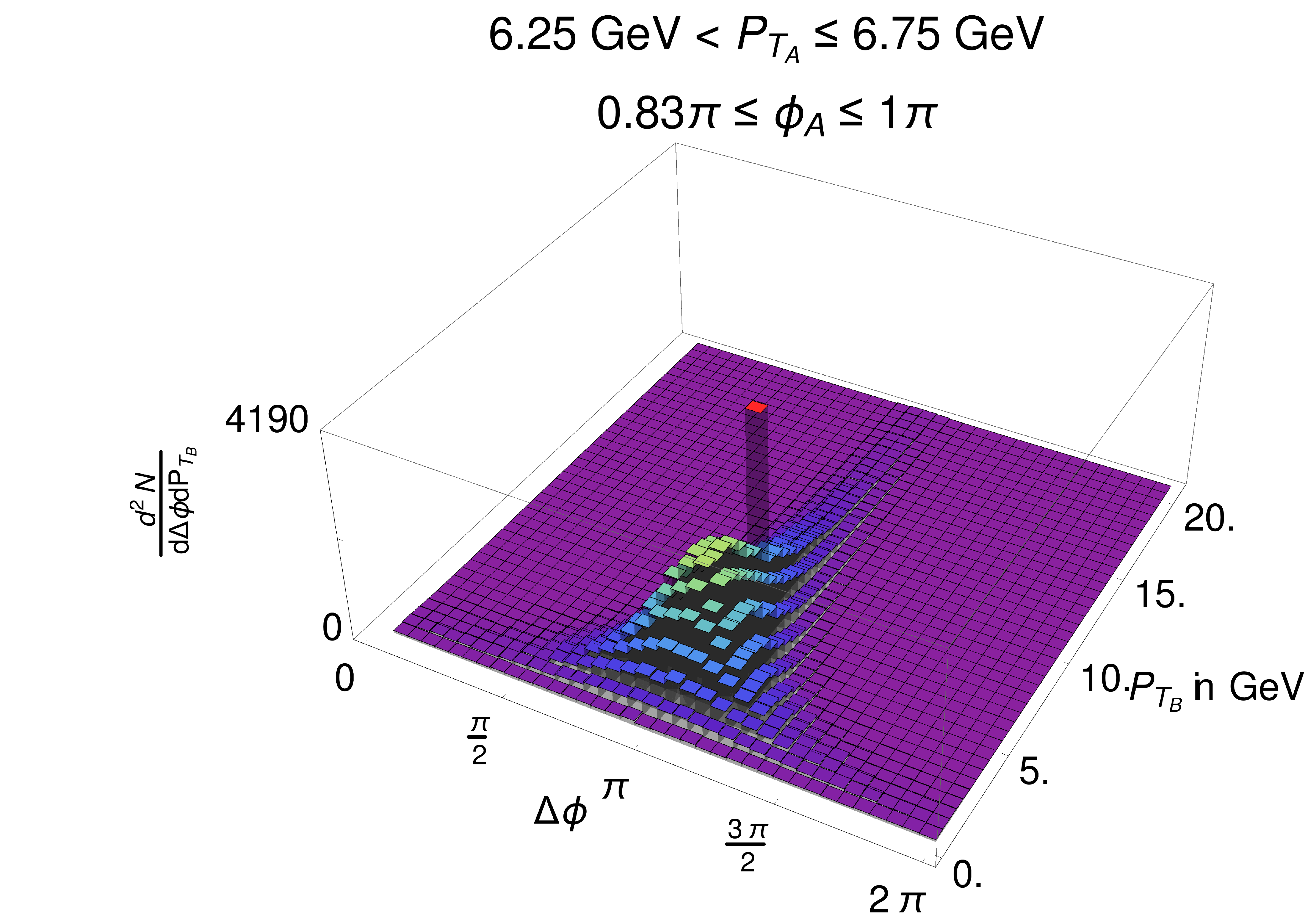}
  }
  \caption{$\frac{d^2N}{d\phi dp_T}$ correlations of $b\bar{b}$ pairs for $p_A=\{2.5, 6.5\}\text{GeV}$ at 40-60\% centrality with $\alpha_s=0.3$}
  \label{fig:2DCorrelationsR}
\end{figure}

  \subsection{2D correlations}
  \label{sec:2D LO correlations}
  In \fg{fig:2DCorrelationsR}, the $\frac{d^2N}{d\phi dp_T}$ correlations are depicted for representative sections of the respective $p_T$ classes.

  We observe that, for low $p_T$, we attain very efficient broadening of the angular correlations. For mid $p_T$, the angular correlations are much tighter, however with greater broadening of the momentum correlations, at least in absolute terms. For $\alpha_s=0.3$, both angular and momentum correlations are typically much weaker than for $\lambda=5.5$, given the larger consequent drag coefficient of the former.

  \subsection{Azimuthal correlations}
  \label{subsec:Azimuthal correlations}
  In \cite{arXiv:1305.3823}, at leading order, the weak coupling based computations exhibited very efficient broadening of initial azimuthal correlations for low $p_T$ $b\bar{b}$ pairs ($[4-10]\text{GeV}$), which were entirely washed out once NLO production processes were taken into consideration.

  Both for mid- and high-$p_T$ ($[4-10]\text{GeV}$ \& $[10-20]\text{GeV}$ respectively), the initial correlations were found to survive to a large degree, both at leading order and at next-to-leading order, suggesting that they may still be observable in an experimental context.
  We compare our strong coupling azimuthal correlations to the weak coupling ones in \fg{fig:AzimuthalCorrelations}.
  For $[10-20\text{GeV}]$, our correlations are significant more peaked at their initial back-to-back correspondence. At $[4-10\text{GeV}]$, this observation still holds for the upper bound of our parameters with $\lambda=5.5$, while the $\alpha_s=0.3$ bounded result is of similar magnitude but looser angular correlation than either the collisional or the collisional + Bremsstrahlung based results. In the $[1-4\text{GeV}]$ range, the azimuthal correlations are almost entirely washed out for $\alpha_s=0.3$, while for $\lambda=5.5$, they are broadened with similiar efficiency to the weak coupling results.

  \begin{figure}
    \centering
    \subfloat{
      \includegraphics[width=.3\textwidth]{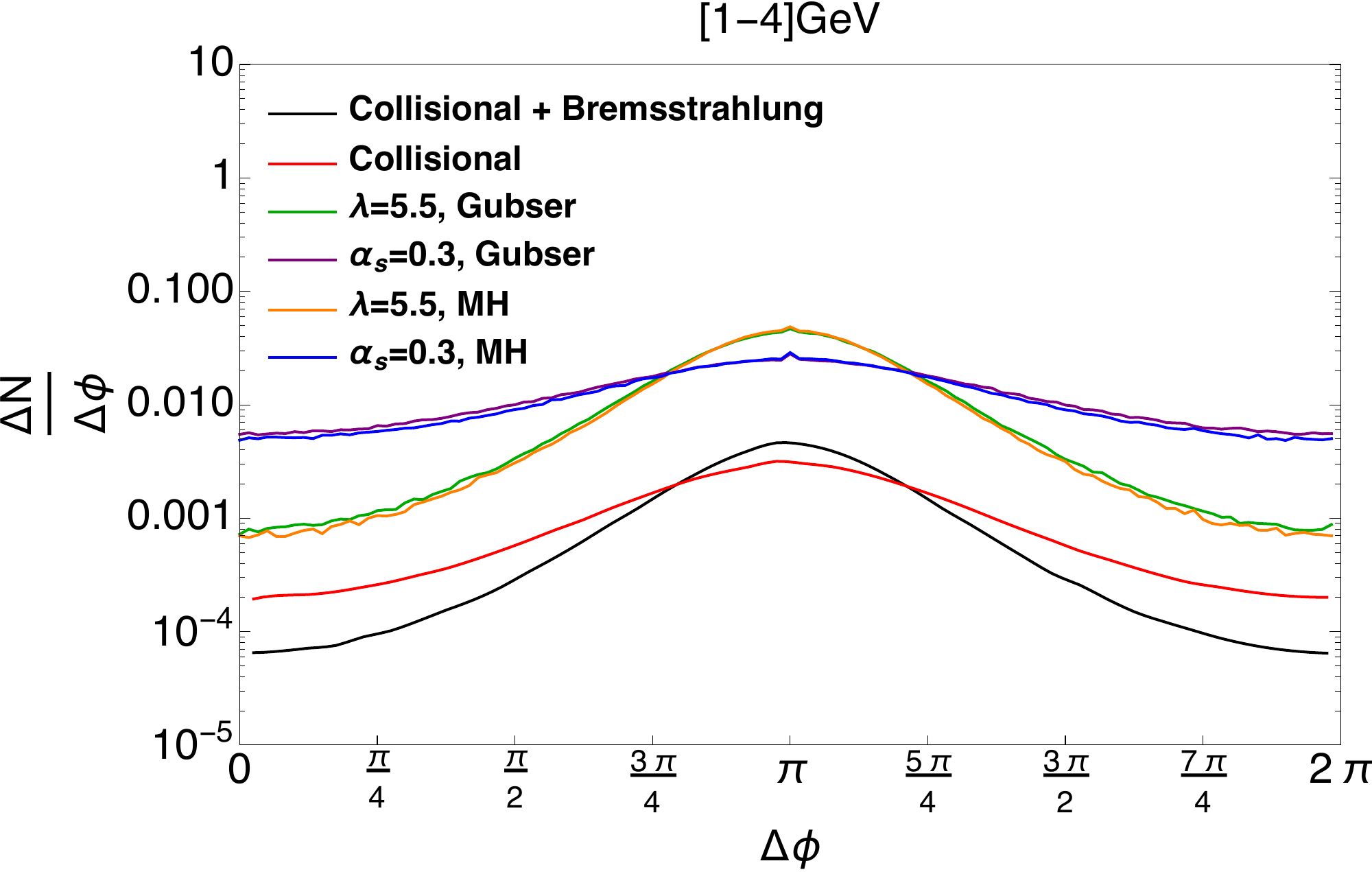}
    }
    \hfill
    \subfloat{
      \includegraphics[width=.3\textwidth]{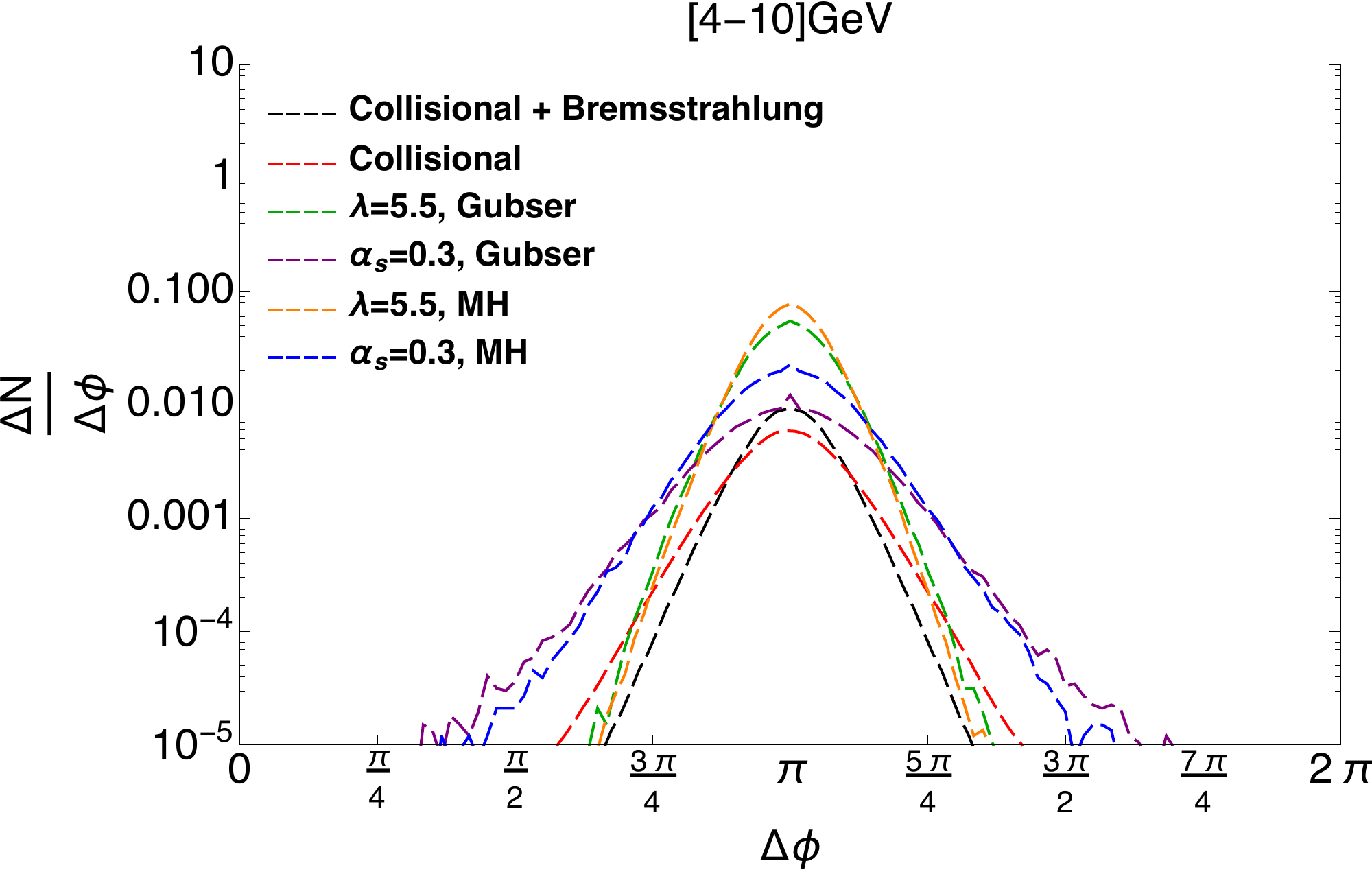}
    }
    \hfill
    \subfloat{
      \includegraphics[width=.3\textwidth]{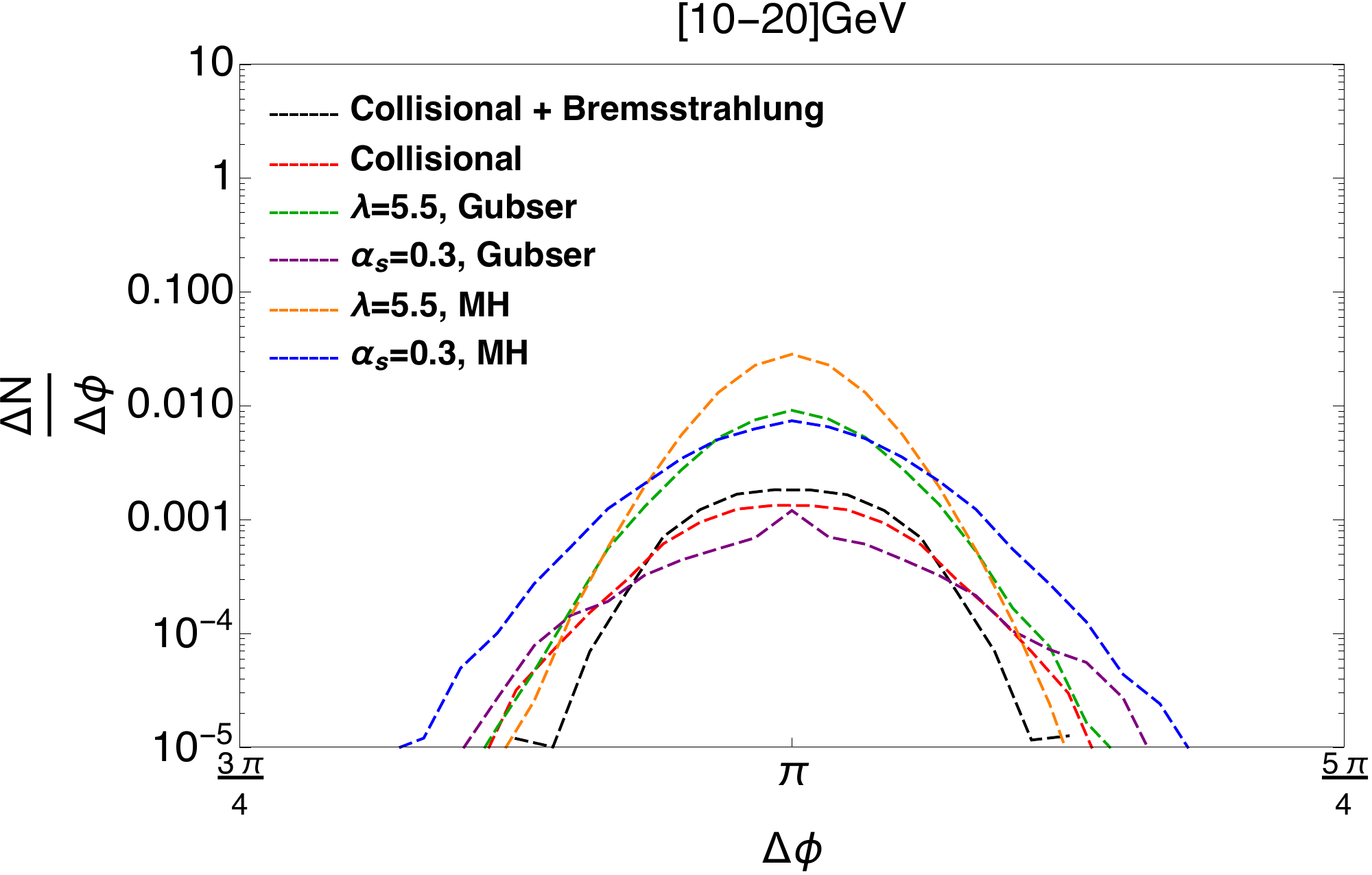}
    }
    \caption{$\frac{dN}{d\phi}$ correlations for the specified classes with 40-60\% centrality.}
    \label{fig:AzimuthalCorrelations}
  \end{figure}

  \subsection{Momentum correlations}
  \label{subsec:Momentum correlations}
  What is striking about the low $p_T$ correlations depicted in \fg{fig:AzimuthalCorrelations} is the massive difference in normalized counts for this momentum class between strong and weak coupling: around an order of magnitude, in fact.

  Na\"{\i}vely, one may expect this to be caused by a more efficient suppression of high momentum particles in a strongly coupled plasma than a weakly coupled one.
  If we take initial momentum correlations into account \fg{fig:Initial Momentum Correlations}, we find that the contribution of particles initially in a higher momentum class being suppressed down to $[1-4]\text{GeV}$ clearly do not suffice to account for the order of magnitude difference observed. In fact, the dominant portion of particles with low final transverse momentum had low transverse momentum initially too.
  On the one hand, this could be due to these particles ending up in other final momentum classes. From \fg{fig:AzimuthalCorrelations}, it is clear they do not end up in $[4-10]\text{GeV}$ or $[10-20]\text{GeV}$. This only leaves $[0-1]\text{GeV}$ and $[20-\infty]\text{GeV}$, the latter being highly unlikely.  

  A more plausible explanation is that in a weakly coupled plasma, $b\bar{b}$ pairs are much more weakly correlated in momentum and are thus more likely to end up in distinct momentum classes.

  Since this effect is observed in the low momentum domain, one may postulate that the momentum fluctuations in a weakly coupled plasma are more significant that in a strongly coupled plasma. Thus, it would be more likely for low momentum heavy quarks to receive momentum kicks that elevate them to higher momentum classes in the weakly coupled plasma.

  \begin{figure}
    \centering
    \includegraphics[width=.359\textwidth]{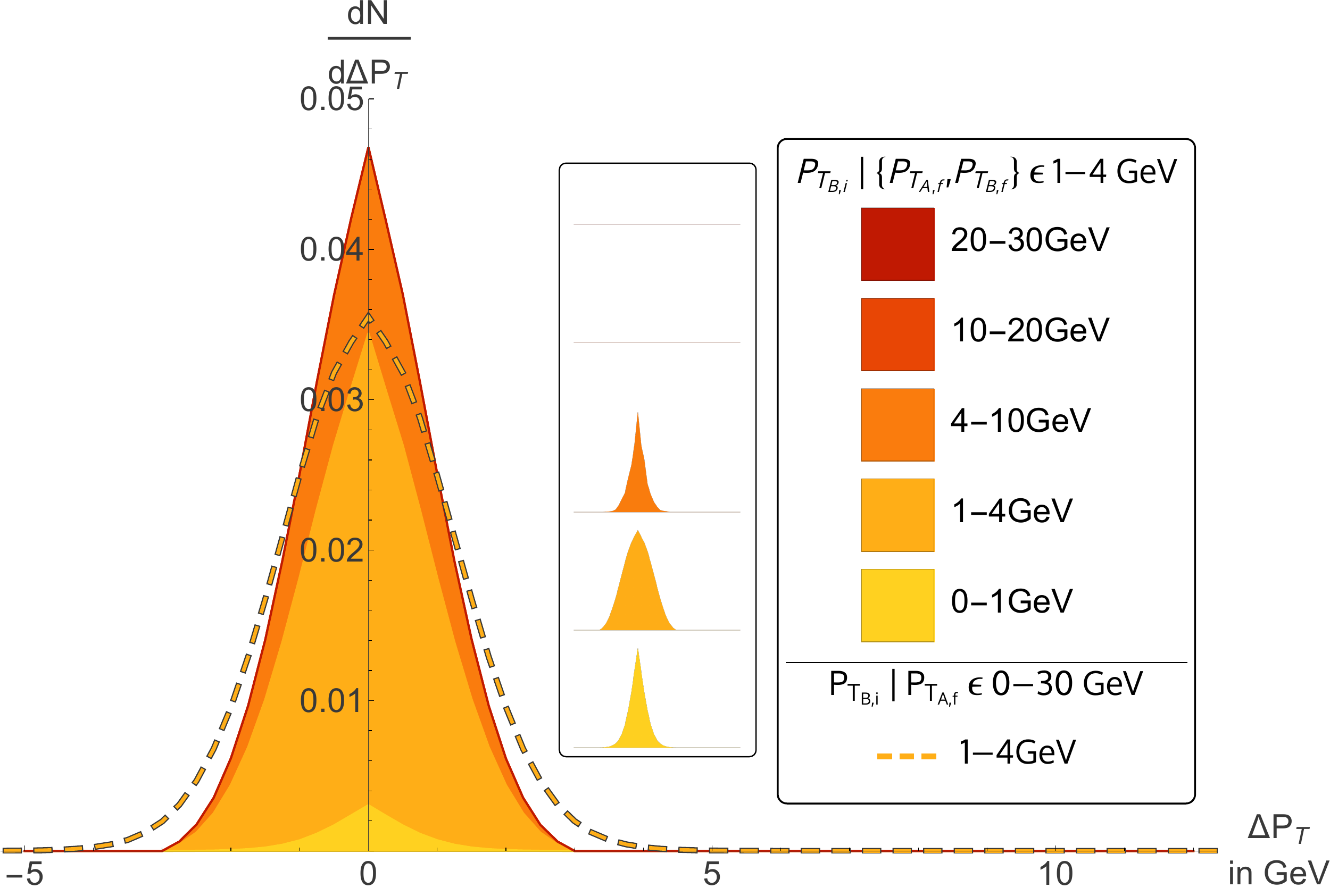}
    \caption{$\frac{dN}{d\Delta p_T}$ of $b\bar{b}$ pairs with $\lambda=5.5$ and initial momentum in some given class.}
    \label{fig:Initial Momentum Correlations}
  \end{figure}

\section{Next-to-leading Order Azimuthal Correlations}
\label{sec:Next-to-leading Order Correlations}


  As depicted in \fg{fig:NLO_final}, the correlations for the $[1-4]\text{ GeV}$ class are entirely washed out, as they've been in \cite{arXiv:1305.3823}. For mid- to high-$p_T$, while the peak around $\Delta\phi=\pi$ has been broadened, the computations suggest that this signal should still be observable in an experimental context. Unfortunately, we had no data from weak coupling calculations for a similar centrality to compare with.

  \begin{figure}
    \centering
    \includegraphics[width=.45\textwidth]{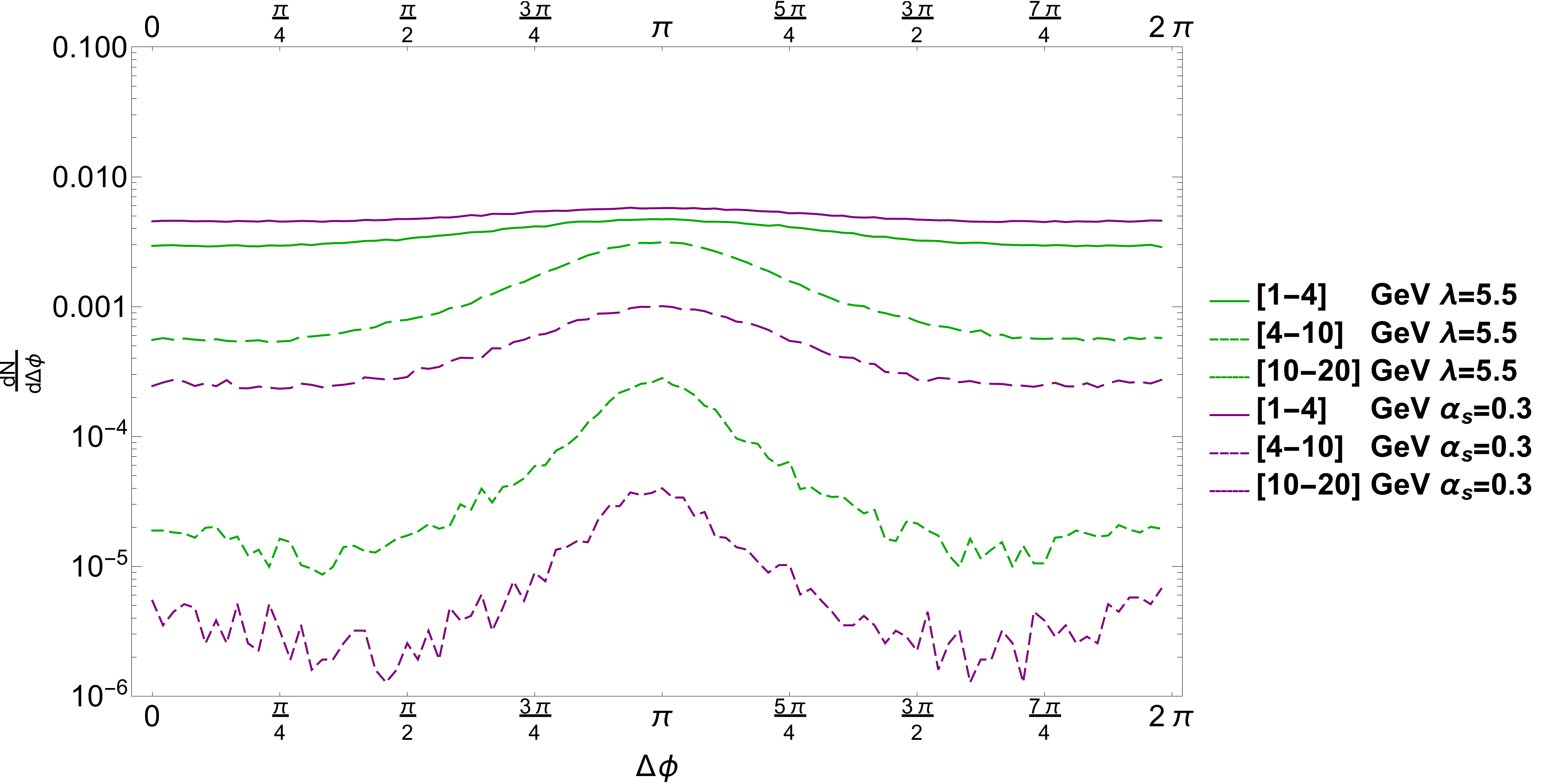}
    \caption{Final $\frac{dN}{d\phi}$ correlations for NLO initialization.}
    \label{fig:NLO_final}
  \end{figure}

\section{Conclusion \& Outlook}
We have compared the azimuthal correlations predicted by pQCD and AdS/CFT based computations and found that, while the azimuthal correlations are qualitatively similar, the momentum correlations tell a different tale.
In particular, the surprise of our findings is the large dissimilarity in low momentum correlations the pQCD and AdS/CFT based simulations exhibit. Thus, bottom quark momentum correlations present an opportunity to distinguish between the energy loss mechanisms of the two frameworks.

Although stronger momentum fluctuations in the weakly coupled plasma have been identified as a plausible explanation for this disparity, we cannot assert this with certainty until initial correlations for the weak coupling based correlations have been considered as well.

Furthermore, whether this order of magnitude difference in predictions for low momentum correlations of heavy quarks exposes weaknesses in either or both of the frameworks cannot be declared until experimental data of bottom quark momentum correlations emerges. Strong coupling based approaches have so far fared better in the low momentum domain, where pQCD is restrained by uncertainties in the running coupling.

While discriminating initial momenta in the momentum correlation plots \fg{fig:Initial Momentum Correlations} is illuminating, producing these is impossible in the current paradigm of detectors, and thus their applicability is restricted to simulations. That being said, the integrated momentum correlations may constitute a potent experimental discriminator of strongly and weakly coupled plasmas and demand further investigation.

\section{Acknowledgments}
The authors wish to thank the South African National Research Foundation (NRF) and the SA-CERN Collaboration for generous support of this work.




\nocite{*}
\bibliographystyle{elsarticle-num}







\end{document}